\documentclass{elsart}

\usepackage{epsf}
\usepackage{epsfig}
\usepackage{psfig}
\usepackage{latexsym}
\usepackage{amssymb}
\usepackage{amsfonts,amsbsy}
\usepackage{amsmath}

\newcommand{\be}{\begin{equation}}
\newcommand{\bel}[1]{\begin{equation}\label{#1}}
\newcommand{\ee}{\end{equation}}
\newcommand{\bea}{\begin{eqnarray}}
\newcommand{\ba}{\begin{array}}
\newcommand{\eea}{\end{eqnarray}}
\newcommand{\ea}{\end{array}}

\begin{document}
\runauthor{Schadschneider}
\begin{frontmatter}
\title{Statistical Physics of Traffic Flow\thanksref{SFB}}
\author[Cologne]{Andreas Schadschneider\thanksref{Email2}}
\thanks[SFB]{Supported by SFB 341 (K\"oln-Aachen-J\"ulich)}

\address[Cologne]{Institut f\"ur Theoretische Physik, Universit\"at zu 
K\"oln, D-50923 K\"oln, Germany}
\thanks[Email2]{E-mail: as@thp.uni-koeln.de}
\begin{abstract} 
The modelling of traffic flow using methods and models from physics
has a long history. In recent years especially cellular automata models
have allowed for large-scale simulations of large traffic networks
faster than real time. On the other hand, these systems are interesting
for physicists since they allow to observe genuine nonequilibrium
effects. Here the current status of cellular automata models for traffic flow
is reviewed with special emphasis on nonequilibrium effects (e.g.\ phase 
transitions) induced by on- and off-ramps.
\end{abstract}
\vspace{0.7cm}
\begin{center}
Physica {\bf A285}, 101 (2000)
\end{center}
\vspace{1cm}
\end{frontmatter}


\section{Introduction}

Despite the long history of application of methods from physics
to traffic flow problems (going back to the fifties) it has only
recently blossomed into a successfull field of ``exotic statistical
physics'' \cite{ourrevs}. Until a few years ago most approaches were based on
"classical" methods from physics, especially from mechanics and
hydrodynamics. In general one can distinguish {\em microscopic}
and {\em macroscopic}  approaches. 

In {\em microscopic models} individual vehicles are distinguished. A typical
example are the so-called {\em car-following theories} \cite{hg,folge}. 
For each car one writes an equation of motion which is the analogue of
Newton's equation. The basic philospophy  of the car-following approach
can be summarized by the equation $[Response]_n =\kappa_n [Stimulus]_n$
for the $n-$th vehicle. Each driver $n$ responds to the surrounding 
traffic conditions which constitute the stimulus for his reaction.
The constant of proportionality $\kappa_n$ is also called sensitivity.
Usually also the reaction-time of the drivers is taken into account.
A typical example is the {\em follow-the-leader} model \cite{folge} where the
stimulus is given by the velocity difference to the next car ahead.
Assuming that drivers tend to move at the same speed as the leading
car the equations of motion are given by $\ddot{x}_{n}(t) = \kappa_n
\left[\dot{x}_{n+1}(t) - \dot{x}_{n}(t)\right]$. In order to obtain
realistic behaviour the sensitivity $\kappa_n$ has to become a function
of the velocity and the distance between the cars, $\kappa_n=
\kappa_n(v_n,x_{n+1}-x_n)$. In recent years the so-called 
{\em optimal-velocity model} \cite{bandoetal} has successfully 
been used. Here the acceleration is determined by the difference of 
the actual velocity $v_n(t)$ and an optimal velocity 
$V^{opt}(\Delta x_n)$ which
depends on the distance $\Delta x_n$ to the next car. The equations
of motion are then of the form $\ddot{x}_{n}(t) = \kappa_n
[V^{opt}(\Delta x_n)-v_n(t)]$. 

In {\em macroscopic models} one does not distinguish individual cars.
Instead a "coarse-grained" fluid-dynamical description in terms of 
densities $c(x,t)$ and flows $J(x,t)$ is used. Traffic is
then viewed as a compressible fluid formed by the vehicles.
Density and flow are related through a continuity equation which
for closed systems takes the form $\partial c /\partial t +
\partial J/\partial x =0$. Since this is only one equation for
two unknown functions one needs additional information.
In one of the first traffic models Lighthill and Whitham \cite{lw} assumed
that $J(x,t)$ is determined by $c(x,t)$, i.e.\ $J(x,t)=J(c(x,t))$.
Inserting this assumption into the continuity equation yields the 
so-called Lighthill-Whitham equation $\partial c /\partial t +
v_g\partial c/\partial x =0$ with $v_g=dJ/dc$. However, for
a more realistic description of an traffic additional equation,
the analogue of the Navier-Stokes equation for fluids, describing
the time-dependence of the velocities $v_n(x,t)$ has to be
considered instead of the simple Lighthill-Whitham assumption.

The first approach borrowing ideas from statistical physics is the
{\em kinetic theory of vehicular traffic} \cite{ph}. Here traffic is 
treated as a gas of interacting particles. The interactions are described
by a dynamical equation in phase space which is the analogue of
the Boltzmann equation in the kinetic theory of gases.

This is only a a very brief overview over the different approaches
for the description of traffic flow. For a more complete exposition
of the models and their merits we refer to the literature, e.g.\
\cite{ourrevs,ph,proc1,proc2,may90,helbook,wolf,kernerPW,nagel99}. 
Nice overviews over the physics of related nonequilibrium models
can be found in \cite{sz,vp,marro,gs}.
In the following we concentrate on a class of models 
developed in the last few years, namely (probabilistic) cellular 
automata. These models describe stochastic processes and are therefore
of general interest for statistical physics \cite{wolfram,droz}.
An extensive review of the application of cellular automata to traffic
flow and related systems can be found in \cite{ourrevs}. Here we focus
on aspects that have not been discussed in much detail in \cite{ourrevs},
e.g.\ the effects of on- and off-ramps.

\section{Some empirical facts} 
\label{secion_empir}

\begin{figure}[ht]
 \centerline{\psfig{figure=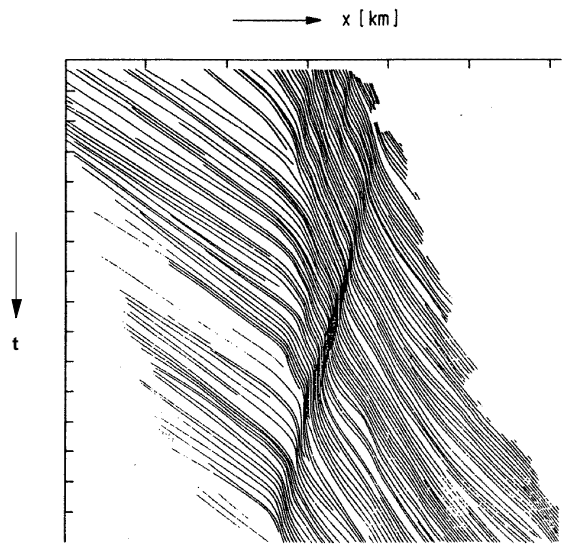,bbllx=14pt,bblly=138pt,bburx=580pt,bbury=700pt,height=8cm}}
  \caption{Trajectories of individual vehicles on a single lane of a 
multi-lane highway showing the spontaneous formation of a jam 
(from \cite{Treiterer75}).}
\label{aerial}
\end{figure}

The most simple empirical facts that should be reproduced by any
traffic model are the spontaneous formation of jams and the
characteristic form of the flow-density relation, the so-called
fundamental diagram. The space-time plot of Fig.~\ref{aerial} shows the 
formation and propagation of a traffic jam. In the beginning, the
vehicles are well separated from each other. Then, without any 
obvious reason like an accident or road construction, a dense region 
appears due to fluctuations. This finally leads to 
the formation of a jam which remains stable for a 
certain period of time but disappears again without any 
obvious reason. 
During its lifetime the front of the jam moves backwards against the
driving direction of the cars with a characteristic velocity of
approximately 15 km/h.

A more detailed analysis of traffic jams in absence of hindrances has
been given by Kerner and Rehborn \cite{kerner,Kerner961,KernerTGF}.  
They found that the upstream velocity and, therefore, the outflow from a
jam is approximately constant. The outflow from a jam and the velocity
of the jam fronts are now regarded as two important empirical
parameters of highway traffic which can be used for calibrating
theoretical models.

\begin{figure}[h]
\begin{center}
\epsfig{file=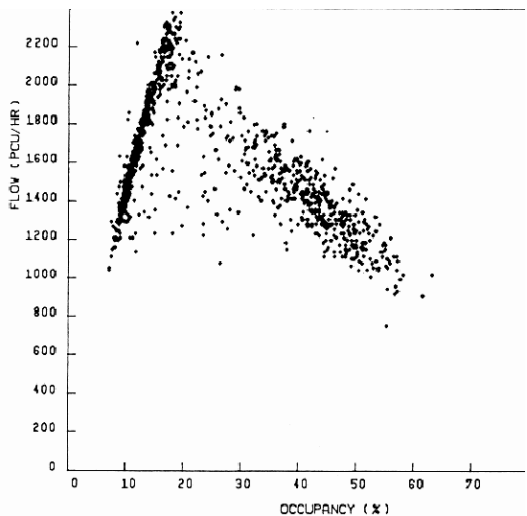,width=7cm}
  \caption{Empirical data for flow and density (occupancy) measured 
    directly by counting loops on a Canadian 
    highway. Each point in the diagram corresponds to an average over a 
    time interval of five minutes (from \cite{Hall86}).}
\label{fig:scatter}
\end{center}
\end{figure}

Fig.~\ref{fig:scatter} shows a typical fundamental diagram obtained
from empirical data. One can clearly distinguish two regions,
the free-flow regime at small densities and the congested regime
at high densities. In the free-flow regime the cars do not hinder
each others motion and therefore the flow grows linearly with 
density. In the congested regime however, the motion of the
vehicles is dominated by the hindrance due to other cars.
Here the flow decreases with increasing density.

These considerations only give the basic structure of the fundamental
diagram. In Sec.~\ref{Sec_gener} a more detailed analysis of empirical data 
will be presented which reveals a much richer structure, e.g.\  the
existance of hysteresis and metastable states. Furthermore in the congested
regime one can distinguish two different phases, namely stop-and-go 
traffic and synchronized flow.


\section{Cellular automata models}
\label{sec_NaSch}

\begin{figure}[ht]
\centerline{\psfig{figure=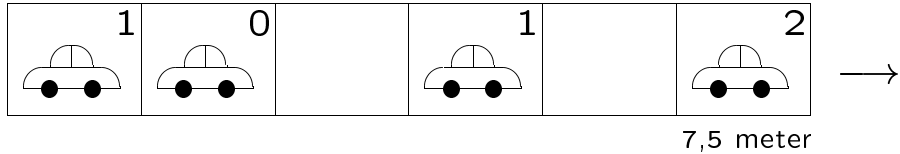,bbllx=50pt,bblly=420pt,bburx=550pt,bbury=540pt,height=3cm}}
\caption{\protect{ A typical configuration in the NaSch model. The 
number in the upper right corner is the speed of the vehicle. }}
\label{carpic}
\end{figure}

The Nagel-Schreckenberg (NaSch) model \cite{ns} is a probabilistic cellular
automaton (CA) able to reproduce the basic features of traffic flow
as discussed in Sec.~\ref{secion_empir}. In a CA not only space and
time are discrete, but also the state variable.
In the NaSch model, each cell can be empty or occupied by exactly
one car (Fig.~\ref{carpic}). 
The state of the car is then characterized by its velocity
$v$ which can take one of the $v_{max}+1$ 
allowed {\it integer} values $v=0,1,...,v_{max}$.
Suppose, $x_n$ and $v_n$ denote the position and velocity, respectively, 
of the $n$-th vehicle. Then, $d_n = x_{n+1}-x_n$, is the gap in 
between the $n$-th car and the car in front of it at time $t$. 
At each time step $t \rightarrow t+1$, the arrangement of the $N$ 
cars on a finite lattice of length $L$ (i.e.\ for a global density
$\rho=N/L$) is updated {\it in parallel} according to the following "rules":

\noindent {\it Step 1: Acceleration.}\\
If $ v_n < v_{max}$, the speed of the $n$-th car is increased by one, i.e.\\
$$
v_n \rightarrow \min(v_n+1,v_{max}) \eqno ({\rm U}1). 
$$ 

\noindent{\it Step 2: Deceleration (due to other cars).}\\
If $d_n \le v_n$, the speed of the $n$-th car is reduced to $d_n-1$, 
i.e.,\\
$$v_n \rightarrow \min(v_n,d_n-1) \eqno ({\rm U}2). $$  

\noindent{\it Step 3: Randomization.}\\
If $v_n > 0$, the speed of the $n$-th car is decreased randomly by 
unity with probability $p$, i.e., \\
$$
v_n \rightarrow \max(v_n-1,0) \quad {\rm with\ probability\ }p 
\eqno ({\rm U}3). 
$$  

\noindent{\it Step 4: Vehicle movement.}\\
Each car is moved forward according to its new velocity determined 
in Steps 1--3, i.e.
$$
x_n \rightarrow  x_n + v_n \eqno ({\rm U}4). 
$$

The NaSch model is a minimal model in the sense that all the four 
steps are necessary to reproduce the basic features of real traffic; 
however, additional rules are needed to capture more complex situations. 
Step 1 reflects the general tendency of the drivers to drive 
as fast as possible without crossing the 
maximum speed limit. Step 2 is intended to avoid collision 
between the cars. The randomization in step 3 takes into 
account the different behavioural patterns of the individual drivers, 
especially, nondeterministic acceleration as well as overreaction 
while slowing down; this is crucially important for the spontaneous 
formation of traffic jams. 
Even changing the precise order of the steps of the update rules stated 
above would change the properties of the model. E.g.\ after changing
the order of steps 2 and 3 there will be no overreactions at braking
and thus no spontaneous formation of jams (see below).
The NaSch model may be regarded as stochastic 
CA~\cite{wolfram}. In the special case $v_{max} = 1$ the deterministic 
limit of the NaSch model is equivalent to the CA rule $184$ in Wolfram's 
notation \cite{wolfram}. 

The use of a {\it parallel} updating scheme (instead of a 
random-sequential one) is crucial \cite{ssni} since it takes into 
account the reaction-time and can lead 
to a chain of overreactions. Suppose, a car slows down due the
randomization step. If the density of cars is large enough this 
might force the following car also to brake in the deceleration 
step. In addition, if $p$ is larger than zero, it might brake even 
further in Step 3. Eventually this can lead to the stopping of a 
car, thus creating a jam. This simple mechanism of spontaneous jam 
formation is rather realistic and cannot be modeled by the 
random-sequential update.

\begin{figure}[ht]
 \centerline{\psfig{figure=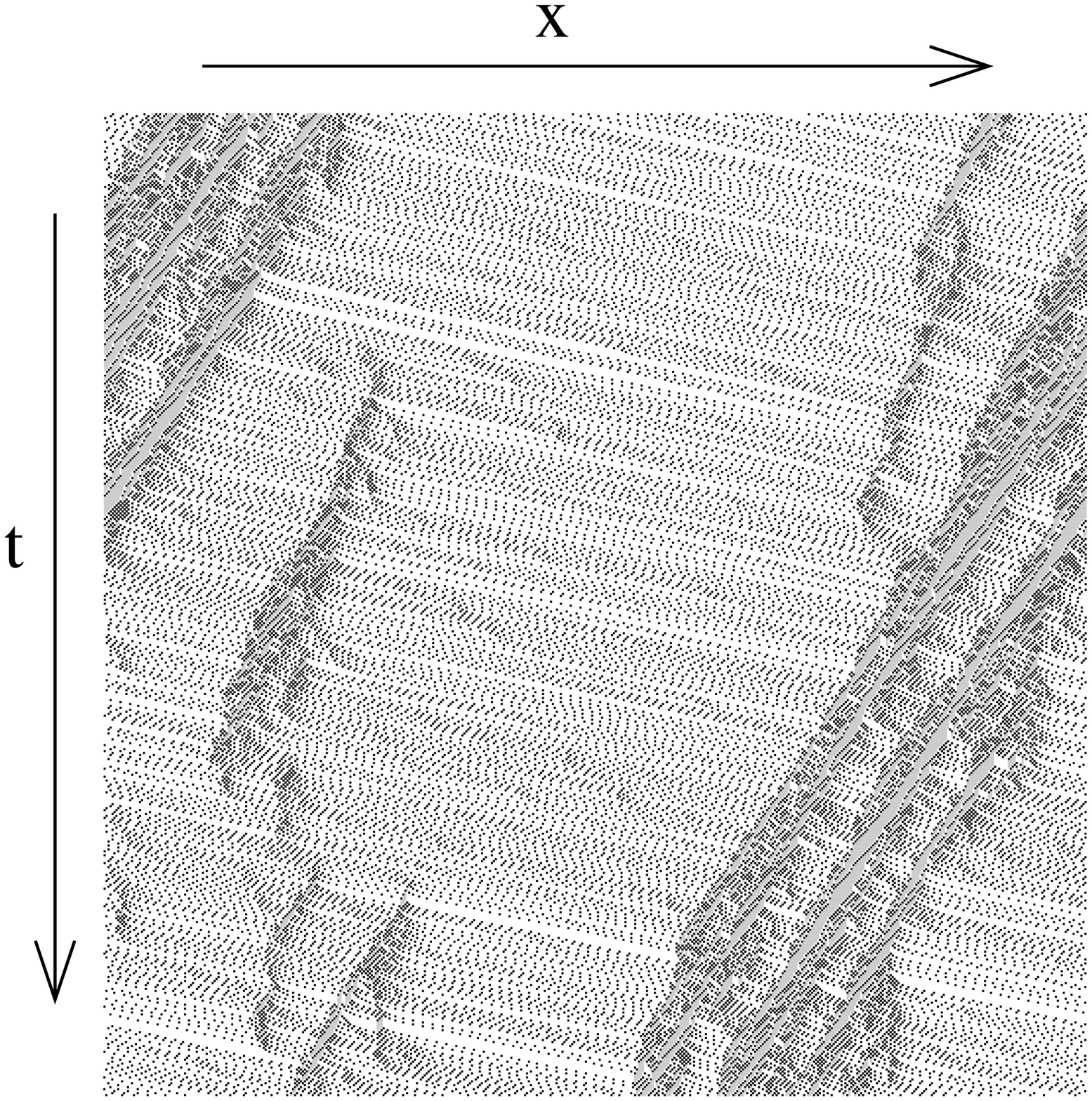,bbllx=10pt,bblly=80pt,bburx=575pt,bbury=680pt,height=6cm}\qquad\qquad
\psfig{figure=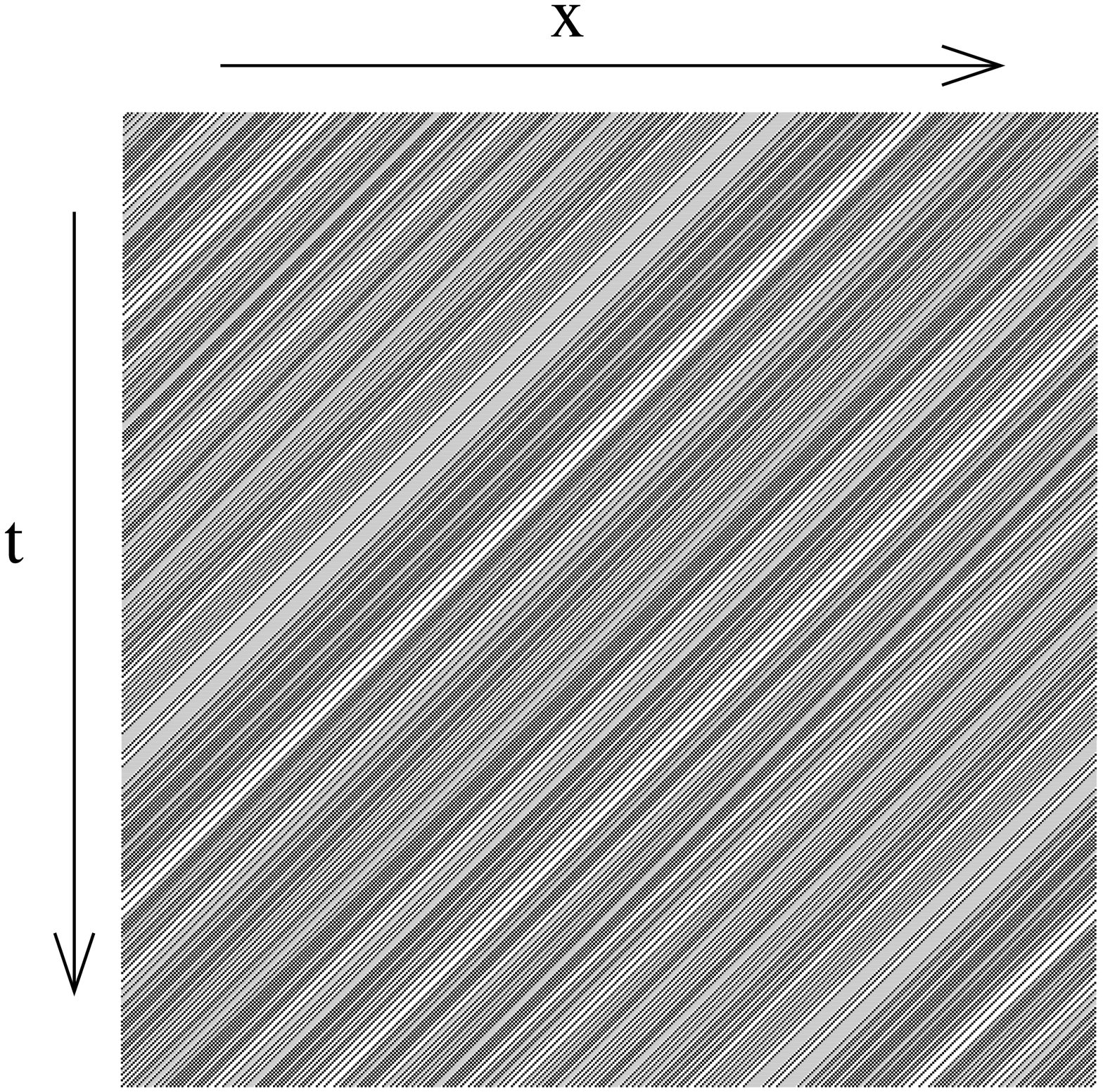,bbllx=10pt,bblly=80pt,bburx=575pt,bbury=680pt,height=6cm}}
\caption{\protect{Typical space-time diagrams of the NaSch model 
with $v_{max} = 5$ and (a) $p = 0.25, \rho = 0.20$, (b) $p = 0, \rho = 0.5$.
}}
\label{nsxt}
\end{figure}

Space-time diagrams showing the time evolutions of the NaSch model 
demonstrate that no jam is present at sufficiently low densities, 
but spontaneous fluctuations give rise to traffic jams at higher 
densities (Fig.~\ref{nsxt}(a)). From Fig.~\ref{nsxt}(b) it should 
be obvious that the {\it intrinsic stochasticity} of the dynamics,
arising from non-zero $p$, is essential for triggering 
the jams. For a realistic description of highway traffic 
\cite{ns}, the typical length of each cell should be about $7.5$~m which
is the space occupied by a car in a dense jam. 
When $v_{max} = 5$ each time step corresponds to approximately 
$1$ sec of real time which is of the order of the shortest relevant
timescale in real traffic, namely the reaction time of the drivers.

\subsection{Open boundary conditions}
\label{openbc}

\begin{figure}[h]
\centerline{\psfig{figure=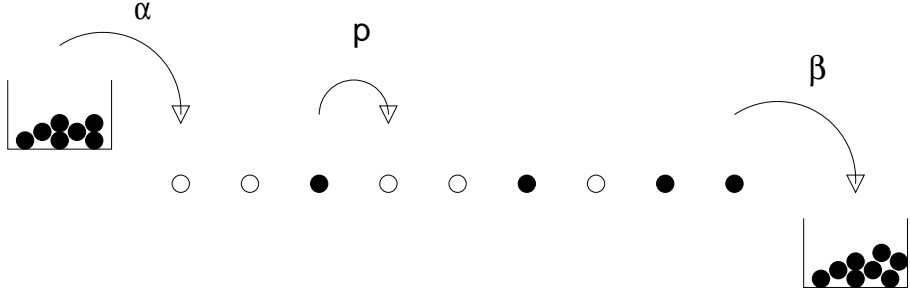,height=4cm}}
\caption{\protect{The asymmetric simple exclusion process.}}
\label{asep}
\end{figure}

For $v_{max}= 1$ the NaSch model reduces to the totally asymmetric simple
exclusion process (ASEP) with parallel update. The ASEP is the simplest 
prototype model of interacting systems driven far from equilibrium 
\cite{sz,gs}.
In the ASEP (Fig.~\ref{asep}) a particle can move forward one cell
with probability\footnote{Note that conventionally
the hopping rate in the ASEP is denoted as $p$. Since in the NaSch
model $p$ is the braking probability the hopping rate in the ASEP (for
$v_{max}=1$) becomes $q=1-p$.}
$p$ if the lattice site immediately in front of it 
is empty. In addition one considers open boundary conditions. If
the first cell is empty a particle will be inserted there with probability
$\alpha$. If the last cell is occupied the particle will be removed
with probability $\beta$.

\begin{figure}[h]
\centerline{\psfig{figure=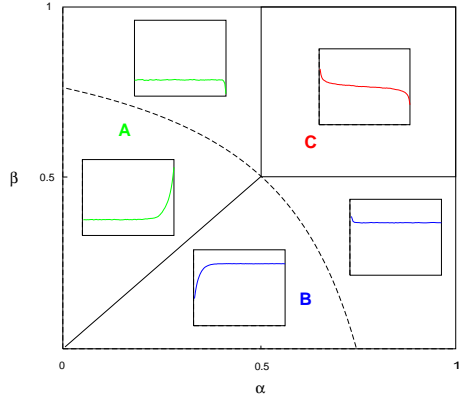,height=5.3cm}}
\caption{\protect{Phase diagram of the ASEP. The inserts show the
typical form of the density profiles.}}
\label{fig_phase}
\end{figure}

By varying the boundary rates $\alpha$ and 
$\beta$ one obtains a surprisingly rich phase diagram (see 
Fig.\ \ref{fig_phase}) which is qualitatively the same for all 
types of dynamics \cite{schdom,derrida93,rsss,ERS,degier}.  
Three phases can be distinguished 
by the functional dependence of the current through the system on
the system parameters. In the low-density phase A ($\alpha < \beta,
\alpha_c(p)$) the current is independent of $\beta$. Here the current
is limited by the rate $\alpha$ which then dominates the
behaviour of the system. In the high-density phase B ($\beta < \alpha,
\beta_c(p)$) the behaviour is dominated by the output rate $\beta$ and 
the current is independent of $\alpha$. In the maximum current phase C
($\alpha>\alpha_c(p)$ and $\beta>\beta_c(p)$) the limiting factor for 
the current is the bulk rate $q=1-p$. Here the current becomes 
independent of both $\alpha$ and $\beta$.

The transitions between the phases can be characterized by the
behaviour of two correlation lengths $\xi_\alpha$ and $\xi_\beta$
which only depend on $p$ and $\alpha$ or $\beta$. 
Apart from $\xi_\alpha$
and $\xi_\beta$ also a third length $\xi^{-1}=|\xi_\alpha^{-1}
-\xi_\beta^{-1}|$ plays an important role.

The transition from A (B) to C is continuous with diverging
correlation length $\xi_\alpha$ ($\xi_\beta$). The transition from the
high- to the low-density phase is of first order. Here both
$\xi_\alpha$ and $\xi_\beta$ are finite, but $\xi$ diverges. On the
transition line one finds a linear density profile created by the
diffusion of a domain wall between a low-density region at the left
end of the chain and a high-density region at the right end.

In \cite{Kolo98} a nice physical picture has been developed which
explains the structure of the phase diagram not only qualitatively,
but also (at least partially) quantitatively.
The phase diagram of the open system is completely determined by
the fundamental diagram of the periodic system through an
extremal-current principle \cite{popkov} and therefore independent
of the microscopic dynamics of the model.

\begin{figure}[h]
\centerline{\psfig{figure=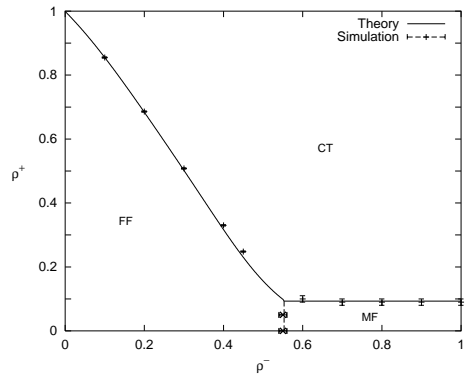,height=5.3cm}}
\caption{\protect{Phase diagram of the NaSch model with open
boundaries for $p=0.25, v_{max}=4$ coupled to reservoirs with
densities $\overline{\rho^{-}}$ and $\overline{\rho^{+}}$
which induce the upstream-density $\rho^-$ and the downstream-density
$\rho^+$.
The phases are: free-flow (FF), congested traffic (CT), maximal flow (MF)
phase.}}
\label{fig_phase5}
\end{figure}

Fig.~\ref{fig_phase5} shows the full phase diagram of the NaSch model
with open boundary conditions and $v_{max}=4$. As predicted by the
extremal-current principle the same phases as in the case $v_{max}=1$
described above appear. Instead of using input- and output-rates
at the boundaries here reservoirs have been introduced which induce
effective densities $\rho^-$ and $\rho^+$ near the boundaries.
In order to allow for an easier comparison with empirical data
obtained near ramps (see Sec.~\ref{ramp_open}) the phases
are denoted as "free-flow" and "congested" instead of "low-density"
and "high-density".

\section{Generalizations and extensions of the NaSch model} 
\label{Sec_gener}

\subsection{Metastability, hysteresis and slow-to-start rules} 
\label{Sec_s2s}

\begin{figure}[ht]
\centerline{\epsfig{figure=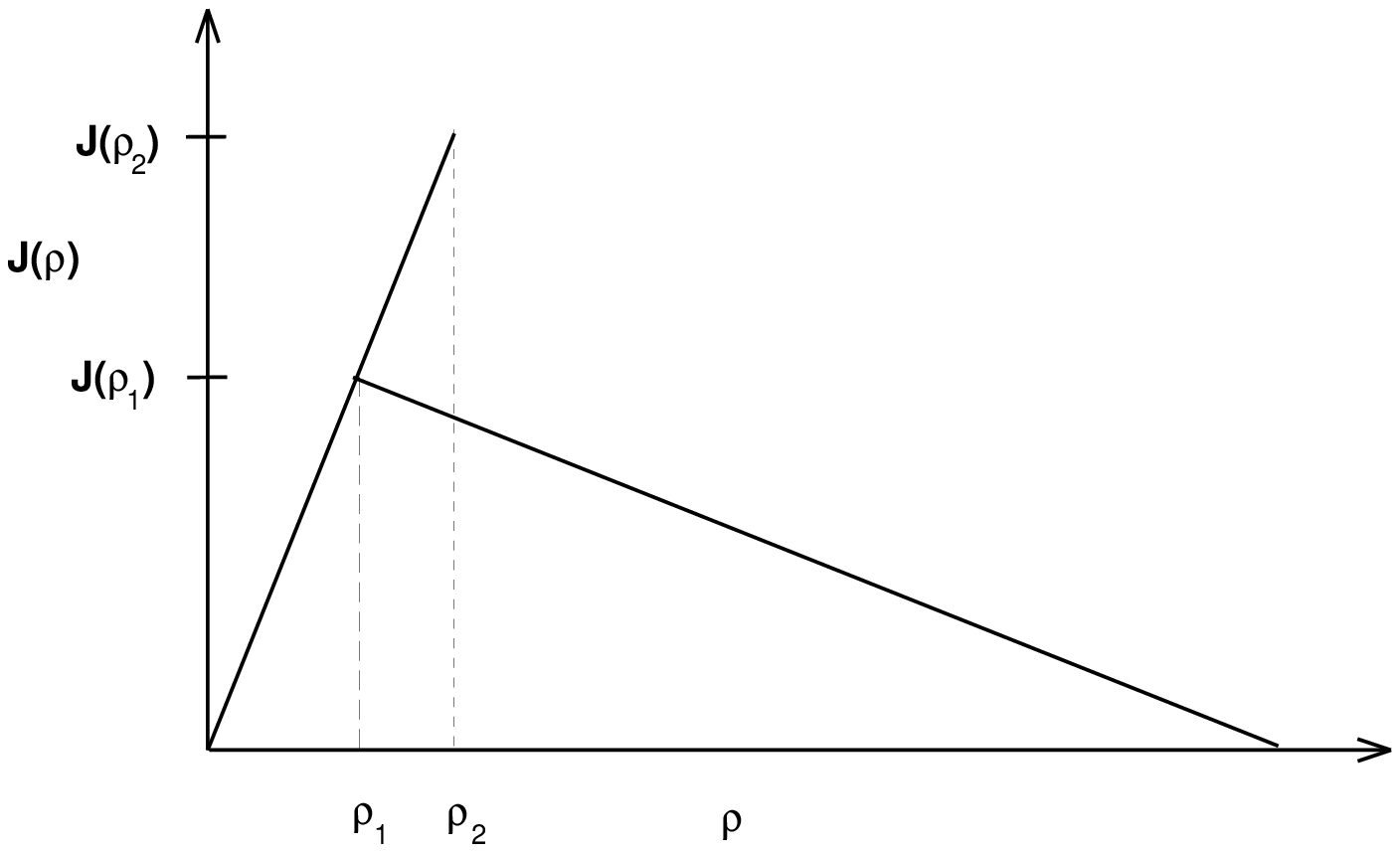,height=5cm}\epsfig{figure=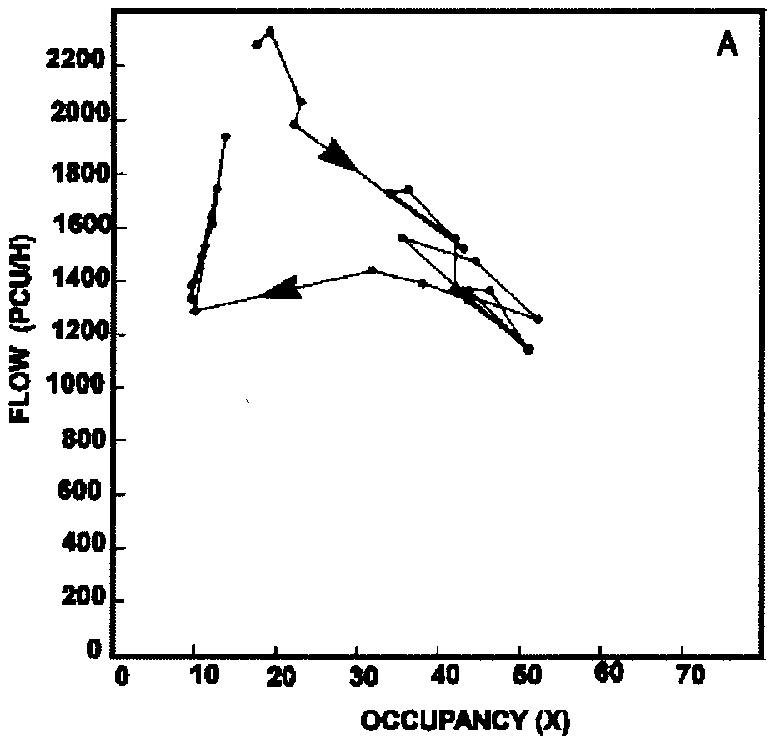,height=6cm}}
\caption{\protect{(a) Schematic form of a fundamental diagram with 
metastable high-flow states, (b) Time-series of empirical flow data
showing hysteresis.}}
\label{metahyst}
\end{figure}

Measurements on real traffic have revealed that traffic flow can exhibit
metastability and hysteresis effects \cite{KernerTGF}. 
Fig.~\ref{metahyst}(a) shows a 
schematic fundamental diagram with a metastable branch.
In the density interval $\rho_1 < \rho < \rho_2$ the flow is not
uniquely determined but depends on the history of the system. Here a
metastable high-flow branch exists. Closely related is the observation
of hysteresis effects. Fig~\ref{metahyst}(b) shows a clear hysteresis loop 
obtained from time-traced measurements of the flow \cite{Hall86}.

The NaSch model in its simplest form as introduced in Sec.~\ref{sec_NaSch}
does not exhibit metastable states and
hysteresis. However, a simple generalization exists which is able to 
reproduce these effects. It is the so-called 
Velocity-Dependent-Randomization (VDR) model \cite{barlovic}.
Here, in contrast to the original NaSch model, the randomization
parameter depends on the velocity of the car, $p=p(v)$.
The rules of Sec.~\ref{sec_NaSch} are supplemented by a new rule,

\noindent {\it Step 0: Determination of the randomization parameter.} 
The randomization parameter used in step 3 for the $n$-th car 
is given by $p=p(v_n(t))$.

This new step has to be carried out before the acceleration step 1.
The randomization parameter used in step 3 depends on the velocity
$v_n(t)$ of the $n-$th car after the previous timestep.
Most relevant for an understanding of the occurance of metastability and
hysteresis are simple slow-to-start rules where one chooses
\cite{barlovic}
\begin{equation}
p(v)=\begin{cases}
p_{0}   & \text{for\ $v = 0$,}\\
p       & \text{for\ $v > 0$,}
\end{cases}
\end{equation}
with $p_0>p$. This means that cars which have been standing in
the previous timestep have a higher probability $p_0$ of braking
in the randomization step than moving cars.

\begin{figure}[ht]
\centerline{\psfig{figure=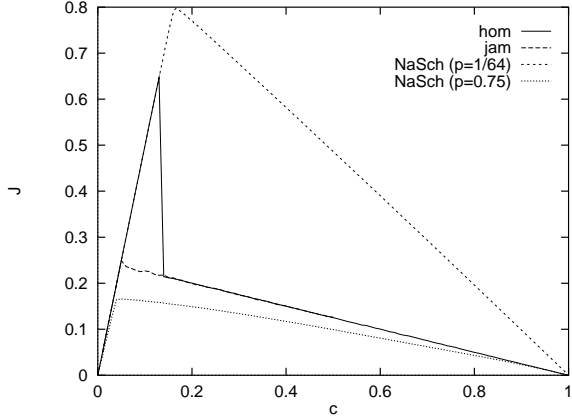,bbllx=50pt,bblly=50pt,bburx=550pt,bbury=400pt,height=5.5cm}}
\caption{\protect{The fundamental diagram in the NaSch model 
with a velocity-dependent slow-to-start rule ($v_{max} = 5$, 
$p_0 = 0.75, p = 1/64$) obtained using two different initial conditions, 
namely, a completely jammed state (jam) and a homogeneous state (hom). }}
\label{vdr_fund}
\end{figure}

\begin{figure}[ht]
 \centerline{\psfig{figure=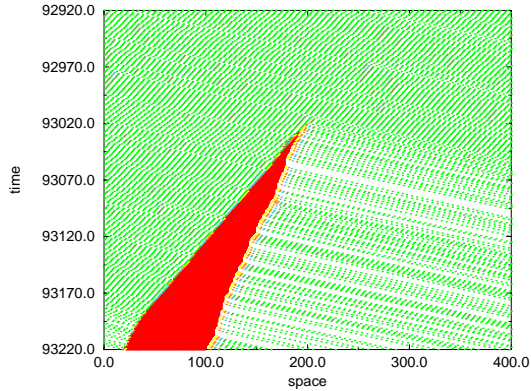,bbllx=12pt,bblly=13pt,bburx=575pt,bbury=480pt,height=6cm}}
\caption{\protect{Typical space-time diagram of the VDR model
    with $v_{max} = 5$ and $\rho = 0.20$, $p = 0.01$ and $p_0 = 0.75$.
    Each horizontal row of dots represents the instantaneous positions
    of the cars moving towards right while the successive rows of
    dots represent the positions of the same cars at the
    successive time steps. }}
\label{xtvdr}
\end{figure}

Fig.~\ref{vdr_fund}  shows a typical fundamental diagram obtained
from computer simulations of the VDR model. It can be clearly seen
that it has the same form as Fig.~\ref{metahyst}(a). 
Over a certain density interval $J(\rho)$ can take one of the two values 
depending on the initial state and, therefore, exhibit metastability.
The microscopic structure of the states in the free-flow branch is
very similar to that of the NaSch model. The metastable homogeneous
states have a lifetime after which their decay leads to congested
branch (see Fig.~\ref{xtvdr}).
In contrast, the microscopic structure of the congested branch differs
drastically from the NaSch model. Here one observes phase separation
into a macroscopic jam and a macroscopic free-flow regime  
(see Fig.~\ref{xtvdr}).

It is instructive to compare the fundamental diagram of the VDR model
with those of the corresponding NaSch models. 
The fundamental diagram of the VDR model can be understood from
heuristic arguments utilizing the observed structures of the
steady-states. For small densities $\rho \ll 1$ there are no slow
cars in the VDR model since interactions between cars are
extremely rare. Every car moves with the free-flow velocity 
$v_{f} = v_{max}- p$ and, therefore, the flux is given by
\begin{equation}
  \label{jhom}
  J_{hom}(\rho) = \rho(v_{max}-p)
\end{equation}
which is identical to the NaSch model with randomization $p$.  On the
other hand, for densities close to $\rho=1$, the cars are likely to
have velocities $v=0$ or $v=1$ only and, therefore, the random braking
is dominated by $p_0$. A simple waiting time argument \cite{barlovic}
shows that the flow in the phase-separated regime is given by
\begin{equation}
  \label{jjam}
  J_{sep}(\rho) = (1-p_0)(1-\rho).
\end{equation}
This corresponds to the NaSch model with randomization $p_0$
(see Fig.~\ref{vdr_fund}).

The behaviour found in the VDR model is generic for models with
slow-to-start rules. Due to the slow-to-start rule the outflow from
a large jam is smaller than the maximal possible flow, i.e.\ the
maximum of the fundamental diagram. Therefore the density far
downstream is rather small and the vehicles propagate almost freely
in the low-density region. The spontaneous formation of
jams is suppressed if $p$ is not too large. This is the basic
mechanism which leads to the formation of the phase-separated state
and hysteresis.


\subsection{Anticipation and synchronized traffic}
\label{sub_synchro}

In the NaSch model as well as in the VDR model
drivers only react to the distance to the next car
ahead, i.e.\ the headway. In real traffic often ``anticipation'' plays
an important role. Drivers not only react to a change of their headway,
but also take into accout the headway of the preceding car. If
this headway is large the preceding car will not brake sharply and
therefore much shorter distances between cars are acceptable.

\begin{figure}
    \begin{center}
    \epsfig{file=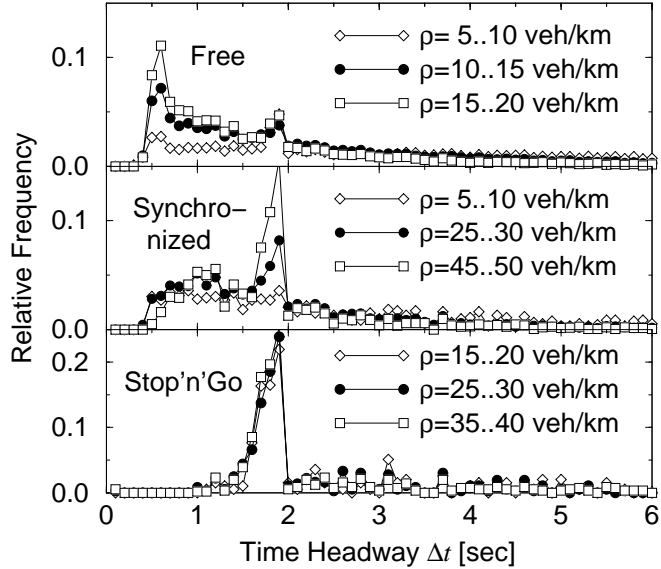,width=0.7\linewidth}
    \caption{Time-headway distribution for different density
      regimes. Top: In free-flow traffic the $\Delta t$-distribution
      is dominated by two peaks at $0.8\sec$ and $1.8\sec$. Middle: In
      synchronized traffic cars with narrow time gaps are to find as
      well as a dominating peak. Bottom: In stop-and-go traffic short
      time headways are surpressed. The peak at $1.8\sec$ remains
      since vehicles are leaving the jam with a typical temporal headway
      of approximately $2\sec$.}
  \label{thfree}
 \end{center}
\end{figure}

Fig.~\ref{thfree} gives an example for this kind of behaviour. It
shows empirical data for time-headways, i.e.\ the time intervals
between the passing of consecutive cars recorded by a detector at
a fixed position on the highway. The time-headway distribution in
the free-flow regime shows a two peak structure. The peak at
$\Delta t =1.8$~sec corresponds to the recommended safe
distance\footnote{This corresponds to the following rule of thumb
taught in (German) driving schools: safe distance (in meters) = half
of velocity (in km/h).}. What is surprising is the existance of a second
even larger peak at $\Delta t=0.8$~sec. On a microscopic level these
short time-headways correspond to platoons
of some vehicles traveling very fast -- their drivers are taking the
risk of driving "bumper-to-bumper" with a rather high speed. These
platoons are the reason for the occurrence of high-flow states in free
traffic. The corresponding states exhibit metastability, i.e. a
perturbation of finite magnitude and duration is able to destroy such
a high-flow state \cite{KernerTGF}.

The NaSch model is not able to reproduce this behaviour. In \cite{newmodel}
a extension of the NaSch model has been introduced which includes 
anticipation as a crucial ingrediant. The driving strategy of the 
drivers in this model comprises four aspects:\\
$(i)$  At large distances the cars move (apart from fluctuations)
with their desired velocity $v_{max}$. 
$(ii)$ At intermediate distances drivers react to velocity changes of 
the next vehicle downstream, i.e.~to ``brake lights''. 
$(iii)$ At small distances the drivers adjust their velocity such that 
safe driving is possible. 
$(iv)$ The acceleration is delayed for standing
vehicles and directly after braking events.
This is achieved by introducing an interaction horizon and brake lights.
The randomization parameter obeys a slow-to-start rule and in addition
depends on the state of the brake light. Furthermore a finer discretization
of the lattice allows for a more realistic modelling of the
acceleration behaviour.

This model is also able to reproduce the typical behaviour found
in so-called synchronized traffic. As first pointed out by Kerner
and Rehborn \cite{kernerPW,Kerner961,Kerner972,Kerner981} 
the congested phase can
be divided into stop-and-go and synchronized traffic. In the latter,
the mean velocity is considerably reduced, but all cars are moving
and the flow is rather large. 

A statistical observable which allows to identify synchronized traffic
is the crosscorrelation
\begin{equation}
cc_{J\rho}=\frac{1}{\sqrt{\sigma(J)\sigma(\rho)}}\left[
  \langle J(t)\rho(t+\tau)\rangle-\langle J(t)\rangle
  \langle\rho(t+\tau)\rangle \right]
\end{equation}
($\sigma(\xi)$ denotes the variance of $\xi$) between the density $\rho$
and the flow $J$ almost vanishes \cite{Neub99}.
This means that these variables are almost independent of each other and
the data points cover randomly a plane in the fundamental diagram
\cite{kernerPW}.
In contrast, in the free-flow phase the flow is completely determined
by the density and $cc_{J\rho}(\tau) \approx 1$.

Another interesting property of synchronized traffic is the existance
of strong velocity correlations between the lanes in multilane 
highways. Lane changes are strongly surpressed and the traffic in
the different lanes becomes ``synchronized''. For a discussion of
further features of synchronized flows and transitions between
the different phases we refer to \cite{kernerPW} (and refs.\ therein).

\section{Effects of on- and off-ramps} 
\label{sec_ramps}

There is large empirical evidence that the activity of on- and off-ramps 
can trigger transitions from free to congested flow
\cite{Neub99,daganzo}. This view was supported by analysis of
macroscopic models, where the introduction of on ramps leads to 
synchronized flow like behaviour \cite{leeetal,hellprl1}.

In a very recent work \cite{DSSZ}, which I will summarize in the 
following subsection, simulations of microscopic models with different 
types of ramps have been discussed.
In Sec.~\ref{ramp_open} experimental evidance \cite{psss}
for the nonequilibrium phase transitions predicted by the ASEP 
(see Sec.~\ref{openbc}) will be presented.

\subsection{Ramps in periodic systems}
\label{ramp_period}
\begin{figure}[ht]
\centerline{\epsfig{figure=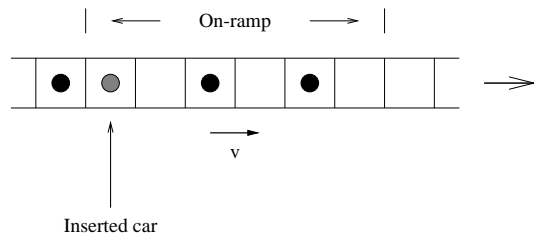,height=3cm}}
\caption{Schematical representation of the on-ramp.}
\label{on_ramp}
\end{figure}

First we discuss implementations of on- and off-ramps in single-lane
systems with periodic boundary conditions where the number of cars 
is constant. The on- and off-ramps are implemented as 
connected parts of the lattice where the vehicles may enter or 
leave the system (see Fig.~\ref{on_ramp}). 
The activity of the ramps is characterised by the number of entering (or
leaving) cars per unit of time  $j_{in}$ ($j_{out}$).  A car is only 
added to the system if the removal of another car at the off-ramp is 
possible at the {\em same} time-step so that $j_{in}=j_{out}$.
In \cite{DSSZ} input and output are performed with a constant
frequency. A stochastic in- and output of cars does 
not lead to a qualitatively different behaviour.
The length chosen for the on- and off-ramps and the distance between
them is motivated by the dimensions found on german highways: $L_{ramp}=25$ 
as length of the ramps in units of the lattice constant (usually 
identified with 7.5 meters). The position, i.e. the first cell of the 
on-ramp, is located at $x_{on}=80$ and that of  
the off-ramp at $x_{off}=L-80$ where $L$ is the system size. Using  
periodic boundary conditions the distance of the on-ramp to the off-ramp is 
given by $d_{ramp}=x_{on}-x_{off}+L$. 

Two different procedures for adding cars to the lattice have been 
investigated in order to model inconsiderate (type A) and 
cautious behaviour (type B) of the drivers at the ramp. The only and essential 
difference in the implementation of the different types of on-ramps lies 
in the strategy of the cars changing from the acceleration lane to 
the driving lane. It has
been found that all the procedures give qualitatively the same results.

\begin{figure}[ht]
\centerline{\epsfig{figure=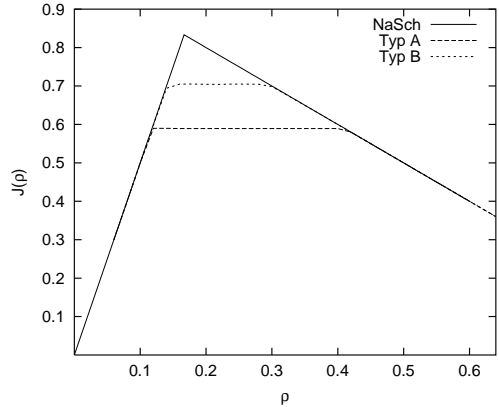,height=5.5cm}}
\caption{Fundamental diagrams of both ramp types at the same input-rate $j_{in}=\frac{1}{5}$. The model parameters are given by $L=3000, v_{max}=5, p=0$.}
\label{fd_ramp}
\end{figure}

In Fig.~\ref{fd_ramp} the fundamental diagrams of the NaSch model for both 
types of on-ramps are shown for the input-rate $j_{in}=j_{out}=\frac{1}{5}$. 
For comparison the fundamental diagram without ramps is also shown 
as solid line.

It is clearly seen that a density regime $\rho_{low}<\rho<\rho_{high}$ 
exists where the flow $J(\rho)$ is independent 
of the density. This so-called plateau value of the flow is lower than the
corresponding flow of the model without ramps. 
An increase of the input-rate $j_{in}$ leads to a decrease of the plateau 
value.
If one compares the different types of input strategies it is evident
that the plateau value of type B is lower than that of  
type A. This can be explained by the probability that the cell in front of 
the inserted car is already occupied. For type B (cautious) this probability 
is smaller than for type A (inconsiderate). 
Nevertheless there is no difference between the two types in a qualitative
sense. 

One can distinguish three different phases depending on the global 
density. In the high and low density phases the 
average flow $J(\rho)$ of the perturbated system takes the same value as in 
the system without ramps. For intermediate densities 
$\rho_{low}<\rho<\rho_{high}$ the flow is constant and limited by the capacity 
of the ramps.

\begin{figure}[ht]
\centerline{\epsfig{figure=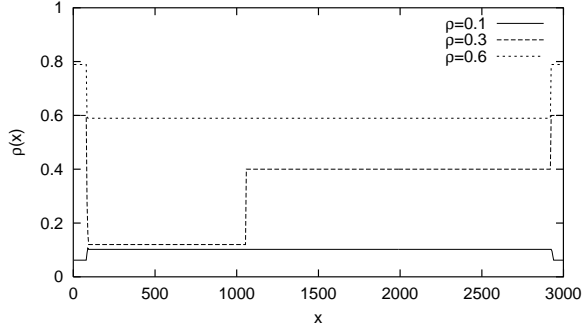,height=4.5cm}}
\caption{Density profiles in the three phases. In the high ($\rho=0.60$) and 
low ($\rho=0.10$) density phases only local inhomogeneities occur near the 
ramps ($x_{on}=80, x_{off}=L-80$). For intermediate densities ($\rho=0.30$) 
one observes phase separation. The model parameters are the same used 
in Fig. \ref{fd_ramp}.}
\label{dp_ramp}
\end{figure}

This behaviour can be understood from the form of the density profiles
(see Fig. \ref{dp_ramp}). In the high and low density phase one only 
observes local deviations from a constant density profile. For intermediate 
densities,however, the system is separated into 
macroscopic high ($\rho_{high}$) and low ($\rho_{low}$) density regions. 
In the region of the ramps the density $\rho_{ramp}$ is additionally higher 
than in the high density region. Thus the ramps act like a blockage in the 
system that decreases the flow locally. Varying the global density within the 
phase-separated state the bulk densities in the high and low density region 
remain constant, only the length of the regions changes. Also the local 
density at the ramps ($\rho_{ramp}$) remains constant. Finally this leads to 
a constant flow in the segregated phase.

This phase separation in systems with ramps and the form of the fundamental
diagram is very similar to the behaviour found in systems with a
stationary defect \cite{lebo,defect,santen}.
Typically, in a certain range on the lattice the slowdown parameter $p$ is 
increased to a value $p_d>p$ compared to that of the residual lattice 
sites \cite{lebo,defect,santen}. 
For intermediate densities 
$\rho_{low}<\rho<\rho_{high}$ the flow is constant in analogy to the model 
with ramps. In this regime $J(\rho)$ is limited by the capacity of the defect 
sites.
There is no qualitative difference between the effect of 
on- and off-ramps and that of stationary defects. In both cases one observes 
plateau formation in the fundamental diagram as well as phase separation in 
the system. The only difference lies in the nature of the blockage dividing
the system into two macroscopic regions. In the case of the ramps it is the 
local increase of the density that decreases the flow locally. In the model
with defects the increased slowdown parameter leads to a local decrease 
of the flow.

\begin{figure}[ht]
\centerline{\epsfig{figure=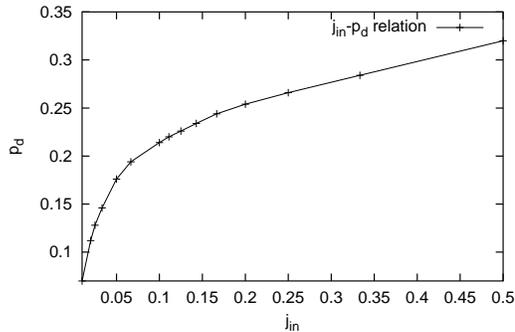,height=4.5cm}}
\caption{Relation between the input-rates $j_{in}$ and the slowdown 
parameter $p_d$ at the defect sites.}
\label{in_pd_dep}
\end{figure}

In order to use the equivalence of ramps and defects in realistic simulations 
of traffic flow one has to determine the relation between the input-rates 
$j_{in}$ and the slowdown parameter $p_d$ at the defect sites. 
Fig. \ref{in_pd_dep} shows the values of $p_d$ and $j_{in}$ leading to the 
same fundamental diagrams. It can be seen that a non-trivial relation between 
the two parameters exists. 

This analogy is very important for efficient
modelling of complex networks \cite{nagel99} with a multitude of
ramps. Therefore it seems to be possible to
achieve a considerable reduction of the computational complexity of
network simulation using this kind of implementation of ramps.

\subsection{Ramps in open systems}
\label{ramp_open}

\begin{figure}[h]
\centerline{\psfig{figure=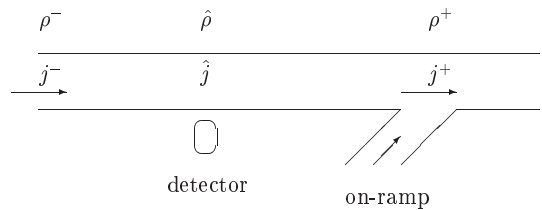,height=3.4cm}}
\caption{\protect{Schematic road design of a highway with an on-ramp
where cars enter the road. The arrows indicate the direction of the flow.
The detector measures the local bulk density $\hat{\rho}$ and bulk flow
$\hat{\jmath}$.}}
\label{fig_ramp}
\end{figure}

In \cite{popkov} it has been shown that empirical data obtained in
measurements on a German highway near Cologne can be interpreted 
using the results presented in Sec.~\ref{openbc}.
These experimental data exhibit boundary effects caused by the presence 
of an on-ramp. Fig.~\ref{fig_ramp} shows a sketch of the relevant part 
of the highway.
No experimental data are available for the density $\rho^-$ and
flow $j^-=j(\rho^-)$ far upstream from the on-ramp and $\rho^+$ and 
$j^+=j(\rho^+)$ just before the on-ramp as well as the activity 
of the ramp. The only data come from a detector located
upstream from the on-ramp which measures a 
traffic density $\hat{\rho}$ and the corresponding flow $\hat{\jmath}$.

Cars entering the motorway cause the mainstream of vehicles
to slow down locally. Therefore, the vehicle density
just before the on-ramp increases to $\rho^{+}>\rho^{-} $.
Then a shock, formed at the on-ramp, will propagate
with mean velocity 
\begin{equation}
v_{shock} = \frac{j(\rho^+) - j(\rho^-)}{\rho^+ - \rho^-},
\label{v_shock}
\end{equation}
obtained from mass conservation. $v_{shock}$ is the velocity of
a 'domain-wall' \footnote{In nonequilibrium systems 
a domain wall is an object connecting two possible stationary states.}
between two stationary regions of densities $\rho^-,\rho^+$
The formation of a stable shock is usually a boundary-driven
process caused by a `bottleneck' on a road. Bottlenecks on a highway
arise from a reduction in the number of lanes and from on-ramps where
additional cars enter a road (see Sec.~\ref{ramp_period}).

Depending on the sign of $v_{shock}$, two scenarios are now possible:\\
1) $v_{shock}>0$ (i.e. $j^{+}> j^{-}$): In this case
the shock propagates (on average) downstream towards the on-ramp. 
Only by fluctuations a brief upstream motion is possible. 
Therefore the detector will measure a traffic density 
$\hat{\rho}=\rho^{-}$ and flow $\hat{\jmath}= j^{-}$.\\
2) $v_{shock}<0$ (i.e. $j^{+}< j^{-}$): 
The shock wave starts propagating with the mean velocity
$v_{shock}$ upstream, thus expanding the congested traffic region with
density $\rho^{+}$. The detector will now measure $\hat{\rho}=\rho^{+}$ 
and flow $\hat{\jmath}= j^{+}$.

In order to discuss the transition between these two scenarios
suppose that one starts with a situation where $j^{+}> j^{-}$
is realized. If now the far-upstream-density $\rho^{-}$ increases 
it will reach a critical point $\rho_{crit}$ above which
$j^{-}> j^{+}$,
i.e., the free flow upstream $j^{-}$ prevails over the flow $j^{+}$ which 
the `bottleneck', i.e. the on-ramp, is able to support.
At this point shock wave velocity $v_{shock}$ will change sign
(see (\ref{v_shock})) and the shock starts traveling upstream.
As a result, the stationary bulk density $\hat{\rho}$ measured by
the detector upstream from the on-ramp will change discontinuously 
from the critical value $\rho_{crit}$ to $\rho^{+}$. This marks a
nonequilibrium phase transition of first order with respect
to order parameter $\hat{\rho}$. The discontinuous change 
of $\hat{\rho}$ leads also an abrupt reduction of the local velocity.
Notice that the flow $\hat{\jmath}=j^{+}$
through the on-ramp (then also measured by the detector) will stay
{\em independent} of the free flow upstream from the congested 
region $j^{-}$ as long as the condition $j^- > j^+$ holds. 

\begin{figure}[h]
\centerline{\psfig{figure=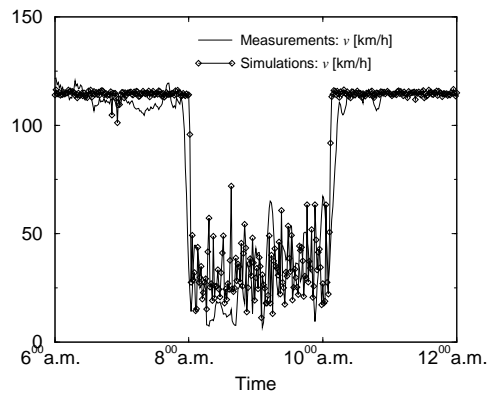,height=5.3cm}}
\caption{\protect{Time series of the velocity. Each data point represents 
an one-minute average of the speed. Shown are empirical data (from 
\cite{Neub99}) in comparison with the results of computer simulations
using a NaSch model with boundary reservoirs.}}
\label{fig_time}
\end{figure}

Empirically this phenomenon can be seen in the traffic data taken from 
measurements on the motorway A1 close to Cologne \cite{Neub99}. 
Fig.~\ref{fig_time} shows a typical time series of the 
one-minute velocity averages. One can clearly see the sharp drop of the 
velocity at about 8 a.m.

\begin{figure}[h]
\centerline{\psfig{figure=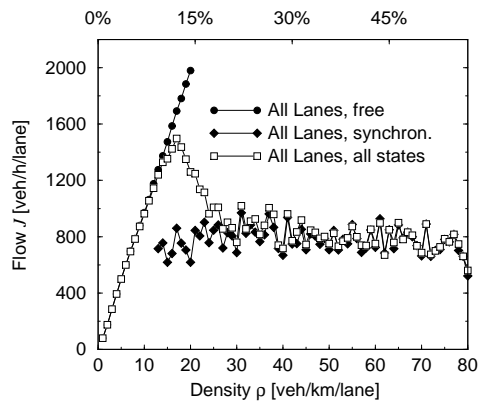,height=5.5cm}}
\caption{\protect{Measurements of the flow versus local density before an
on-ramp on the motorway A1 close to Cologne. The detector is located at a
distance 1 km upstream from on-ramp.}}
\label{fig_measur}
\end{figure}

Also the measurements of the flow versus local density,
i.e.\ the fundamental diagram (Fig.~\ref{fig_measur}), support this
interpretation. Two distinct branches can be distinguished. 
The increasing part corresponds to an almost linear rise of the flow 
with density in the free-flow regime. This part of the flow diagram is 
not affected by the presence of the on-ramp at all and one 
measures $\hat{\jmath} = j^{-}$ which is the actual upstream flow. 
The second branch are measurements taken during congested traffic hours.
The transition from free flow to congested traffic is characterized 
by a discontinuous
reduction of the local velocity. However, as predicted above
the flow does not change significantly in the congested regime.
In contrast, in local measurements large density fluctuations can
be observed. Therefore in this regime the density does not take the 
constant value $\rho^{+}$ as suggested by the argument given above,
but varies from 20 veh/km/lane to 80 veh/km/lane (see Fig.~\ref{fig_measur}).
In contrast, spontaneously emerging and decaying jams would lead to the
observation of a non-constant flow.

This shows that the empirical data obtained near an on-ramp can indeed
be interpreted as a nonequilibrium phase transition. Furthermore
it shows exactly the same behaviour as the transition from the
low-density to the high-density phase in the ASEP (see Sec.~\ref{openbc}).

The case of an off-ramp (or expansion of road space etc.) leads to a 
local decrease $\rho^+ <\rho^-$ of the density. Here the collective 
velocity $v_c= \frac{\partial  j(\rho) }{\partial \rho} $ plays an
important role. The theory than predicts a transition from the free-flow
regime to the maximal flow (MF) phase \cite{psss}. This transition 
is of second order because $\hat{\rho}$ changes continuously across 
the phase transition point and has not yet been observed empirically.
The existence of a maximal flow phase was not emphasized in the
context of traffic flow up to now. At the same time, it is the most
desirable phase, carrying the maximal possible throughput of vehicles
$j_{max}$.  

For practical purposes above observations may directly be used in
order to operate a highway in the optimal regime. E.g.\ the flow near a
lane reduction could be increased significantly if the traffic state
at the entry would allow to attain the maximal possible flow 
of the bottleneck. This could be achieved by controlling the density 
far upstream, e.g.\ by closing temporally an on-ramp, such 
that the cars still enter the bottleneck with high velocity.


\section{Summary and conclusion} 
\label{Sec_summ}

In this short review only a few examples for the successfull
application of CA models for the description of traffic flow
have been presented. Despite its conceptual simplicity the NaSch model 
and its variants are able to reproduce many phenomena observed
empirically. Although the same is true for many of the other
approaches mentioned here only briefly in the introduction CA
have the big advantage of being ideally suited for large-scale
computer simulations. It is now possible to simulate large
networks of highway \cite{ZPR} and city traffic 
\cite{duisburg,dallas,geneva}
faster than real time. This offers new perspectives for the 
planning and design of transportation networks. Finally it is
worthwhile pointing out that not only traffic science profits
from the application of methods from physics. The example of
the first empirical observation of a boundary-induced phase transition
(see Sec.~\ref{ramp_open}) shows that also physics can profit
from applications to "exotic fields".


\vspace{2cm}

\noindent{\bf Acknowledgements:} I like to thank all my collaborators,
especially L.\ Santen and D.\ Chowdhury for their contribution to
the extensive review \cite{ourrevs}.
I also like to thank the organizers of the 36.\ Karpacz Winter School
for creating a stimulating atmosphere.

\vspace{1cm}

\newpage 




\begin{thebibliography}{999}

\bibitem{ourrevs} D. Chowdhury, L. Santen, A. Schadschneider, 
Phys.\ Rep.\ {\bf 329}, 199 (2000) and Curr. Sci. {\bf 77}, 411 (1999)

\bibitem{hg} R. Herman and K. Gardels, 
 Sci. Am. {\bf 209}(6), 35 (1963)

\bibitem{folge} L.A.\ Pipes, J.\ Appl.\ Phys.\ {\bf 24}, 274 (1953)

\bibitem{bandoetal} M. Bando, K. Hasebe, A. Nakayama, A. Shibata and
  Y. Sugiyama, Phys. Rev. E {\bf 51}, 1035 (1995); 
  Jpn. J. Ind. Appl. Math. {\bf 11}, 202 (1994)

\bibitem{lw} M.J. Lighthill and G.B. Whitham, Proc. Roy. Soc. Lond. A
{\bf 229}, 281 (1955)

\bibitem{ph} I. Prigogine and R. Herman, {\em Kinetic Theory of
    Vehicular Traffic} (Elsevier, Amsterdam, 1971)

\bibitem{proc1} D.E. Wolf, M. Schreckenberg and A. Bachem (eds.), {\em
    Traffic and Granular Flow} (World Scientific, Singapore, 1996)

\bibitem{proc2} M. Schreckenberg and D.E. Wolf (eds.), {\em Traffic and
    Granular Flow '97} (Springer, Singapore, 1998)

\bibitem{may90} A.D. May, {\em Traffic Flow Fundamentals} 
(Prentice-Hall, 1990)

\bibitem{helbook} D. Helbing, {\em Verkehrsdynamik: Neue Physikalische
    Modellierungskonzepte} (in German) (Springer, 1997) 

\bibitem{wolf} D.E. Wolf, Physica A {\bf 263}, 438 (1999)

\bibitem{kernerPW} B.S. Kerner, Phys. World {\bf 8}, 25 (1999)

\bibitem{nagel99} K. Nagel, J. Esser and M. Rickert, in: 
Annu. Rev. Comp. Phys. 7, p. 151, ed. D. Stauffer (World Scientific, 2000)

\bibitem{sz} B. Schmittmann and R.K.P. Zia, in: {\em Phase Transitions
    and Critical Phenomena}, Vol.~17, eds. C. Domb and J.L. Lebowitz 
  (Academic Press, 1995)

\bibitem{vp} V. Privman (ed.), {\em Nonequilibrium Statistical
  Mechanics in One Dimension} (Cambridge University Press, 1997)

\bibitem{marro} J. Marro and R. Dickman, {\em Nonequilibrium Phase 
Transitions in Lattice Models} (Cambridge University Press, 1999)

\bibitem{gs} G.M. Sch\"utz, in: {\em Phase Transitions and Critical
    Phenomena}, Vol.\ 19, eds. C. Domb and J.L. Lebowitz 
    (Academic Press, 2000)

\bibitem{wolfram} S. Wolfram, {\em Theory and Applications of Cellular
    Automata}, (World Scientific, 1986); {\em Cellular Automata and
    Complexity} (Addison-Wesley, 1994)

\bibitem{droz} B. Chopard and M. Droz, {\em Cellular Automata
    Modelling of Physical Systems} (Cambridge University Press, 1998)

\bibitem{Treiterer75} J.\ Treiterer: Ohio State Technical Report 
No.\ PB 246 094, (1975)

\bibitem{kerner} B.S. Kerner and H. Rehborn, Phys. Rev. E {\bf 53},
  R1297 (1996)

\bibitem{Kerner961}  B.S. Kerner and H. Rehborn,
  Phys. Rev. E {\bf 53}, R4275 (1996)

\bibitem{KernerTGF} B.S. Kerner, in \cite{proc2}, p. 239

\bibitem{Hall86} F.L. Hall, B.L. Allen and  M.A. Gunter,
  Transp. Res. {\bf A20}, 197 (1986)

\bibitem{ns} K. Nagel and M. Schreckenberg, 
J. Physique I, {\bf 2}, 2221 (1992)

\bibitem{ssni} M. Schreckenberg, A. Schadschneider, K. Nagel and N. Ito, 
Phys. Rev. E {\bf 51}, 2939 (1995)

\bibitem{schdom} G.\ Sch\"utz and E.\ Domany, 
J. Stat. Phys.{\bf 72}, 277 (1993)

\bibitem{derrida93} B.\ Derrida, M.R.\ Evans, V.\ Hakim and
V.\ Pasquier, 
J. Phys. A {\bf 26}, 1493 (1993)

\bibitem{rsss} N.\ Rajewsky, L.\ Santen, A.\ Schadschneider and 
M.\ Schreckenberg, J.\ Stat.\ Phys.\ {\bf 92}, 151 (1998)

\bibitem{ERS} M.R.\ Evans, N.\ Rajewsky and E.R.\ Speer,
J. Stat. Phys. {\bf 95}, 45 (1999) 

\bibitem{degier} J.\ de Gier and B.\ Nienhuis, 
Phys. Rev. E {\bf 59}, 4899 (1999)

\bibitem{Kolo98} A.B.\ Kolomeisky, G.~Sch\"utz, E.B.\ Kolomeisky
and J.P.\ Straley, J. Phys. A {\bf 31}, 6911 (1998)

\bibitem{popkov} V. Popkov and G. Sch\"utz, 
Europhys.\ Lett. {\bf 48}, 257 (1999)

\bibitem{barlovic} R. Barlovic, L. Santen, A. Schadschneider 
and M. Schreckenberg, Eur. Phys. J. B {\bf 5}, 793 (1998)

\bibitem{newmodel} W. Knospe, L. Santen, A. Schadschneider and
M. Schreckenberg, preprint (2000)
 
\bibitem{Kerner972}  B.S. Kerner and H. Rehborn,
  Phys. Rev. Lett. {\bf 79}, 4030 (1998)

\bibitem{Kerner981}  B.S. Kerner,
  Phys. Rev. Lett. {\bf 81}, 3797 (1998)

\bibitem{Neub99} L. Neubert, L. Santen, A. Schadschneider and M.
  Schreckenberg, Phys. Rev. E {\bf 60}, 6480 (1999)

\bibitem{daganzo} C.F. Daganzo, M.J. Cassidy, R.L. Bertini,
  Transp.~Res.~A {\bf 33}, 365 (1999)

\bibitem{leeetal} H.Y. Lee, H.-W. Lee, D. Kim, Phys. Rev. Lett.
  {\bf 81}, 1130 (1998); Phys. Rev. E {\bf 59}, 5101 (1999)

\bibitem{hellprl1} D.~Helbing,  M.~Treiber, 
                   Phys.~Rev.~Lett. {\bf 81}, 3042 (1998)

\bibitem{DSSZ} G. Diedrich, L. Santen, A. Schadschneider and
J. Zittartz, Int. J. Mod. Phys. {\bf C11}, 335 (2000)

\bibitem{psss} V. Popkov, L. Santen, A. Schadschneider
and G.M. Sch\"utz, {\tt cond-mat/0002169}

\bibitem{lebo} S.A. Janowsky and J.L. Lebowitz, 
Phys. Rev. A {\bf 45}, 618 (1992); J. Stat. Phys. {\bf 77}, 35 (1994) 

\bibitem{defect} W. Knospe, L. Santen, A. Schadschneider, 
M. Schreckenberg, in \cite{proc2}, p. 349 (and refs.\ therein)

\bibitem{santen} L. Santen, {\em Numerical Investigations of 
Discrete Models for Traffic Flow}, Ph.D. thesis, Universit\"at zu 
K\"oln (1999)

\bibitem{ZPR} M. Rickert, Diploma thesis, Universit\"at zu K\"oln (1994);
K. Nagel, Ph.D. thesis, Universit\"at zu K\"oln (1995)

\bibitem{duisburg} J. Esser and M. Schreckenberg, 
Int. J. Mod. Phys.C {\bf 8}, 1025 (1997).

\bibitem{dallas} M. Rickert and K. Nagel, 
Int. J. Mod. Phys.C {\bf 8}, 483 (1997); 
K. Nagel and C.L. Barrett, Int. J. Mod. Phys. C {\bf 8}, 505 (1997)

\bibitem{geneva} B. Chopard, A. Dupuis and P. Luthi, 
in ref.\ \cite{proc2}, p. 153


\end{thebibliography}
\end{document}